\documentclass[usenatbib]{mnras}

\usepackage{newtxtext,newtxmath}

\usepackage[T1]{fontenc}
\usepackage{ae,aecompl}

\usepackage{graphicx}	
\usepackage{amsmath}	
\usepackage{amssymb}	
\usepackage{siunitx}

\newcommand{\dss}{$\delta$~Sct stars}
\newcommand{\ds}{$\delta$~Sct}
\newcommand{\gdor}{$\gamma$~Dor}
\newcommand{\ep}{$\epsilon$}
\newcommand{\Ga}{$\Gamma{_j}$}

\title[Self-consistent method to extract non-linearities]{Self-consistent method to extract non-linearities from pulsating stars light curves I. Combination frequencies. }

\author[Lares-Martiz, M. et al.]{
Lares-Martiz ,M.,$^{1}$\thanks{E-mail: mlm@iaa.es }
Garrido, R.,$^{1}$
Pascual-Granado, J.$^{1}$
\\
$^{1}$Instituto de Astrof\'isica de Andaluc\'ia (IAA-CSIC). Glorieta de Astronom\'ia s/n, 18008, Granada. Spain \\
}

\date{Accepted XXX. Received YYY; in original form ZZZ}

\pubyear{2020}

\begin{document}
\label{firstpage}
\pagerange{\pageref{firstpage}--\pageref{lastpage}}
\maketitle

\begin{abstract}

Combination frequencies are not solutions of the perturbed stellar structure equations. In dense power spectra from a light curve of a given multi-periodic pulsating star, they can compromise the mode identification in an asteroseismic analysis, hence they must be treated as spurious frequencies and conveniently removed. In this paper, a method based on fitting the set of frequencies that best describe a general non-linear model, like the Volterra series, is presented. The method allows to extract these frequencies from the power spectrum, so helping to improve the frequency analysis enabling hidden frequencies to emerge from the initially considered as noise. Moreover, the method yields frequencies with uncertainties several orders of magnitude smaller than the Rayleigh dispersion, usually taken as the present error in a standard frequency analysis. Furthermore, it is compatible with the classical counting cycles method, the so-called O-C method, which is valid only for mono-periodic stars. The method opens the possibility to characterise the non-linear behaviour of a given pulsating star by studying in detail the complex generalised transfer functions.

\end{abstract}




\section{Introduction}

Combination frequencies arise in the power spectra of some pulsating stars (\dss, \gdor~stars, Cepheids, $\beta$ Cep stars, SPB stars, RR lyrae stars or white dwarfs) due to a non-sinusoidal shape of their light curves.

Such deviations may have physical explanations related to the non-linear response of the stellar interior to the oscillation wave, as well as for mechanisms that follow non-linear dependencies (e.g. the $T^4$ dependency in the Stefan-Boltzmann law for the emergent flux). Their consequences are non-linear interactions of the intrinsic pulsation modes of the star, generating cross-terms at summation, difference and harmonics of them. 

These were analysed by different authors: \citet{Fokin} studied the propagation of a shockwave in RR Lyrae stars; or \citet{1992MNRAS.259..519B} for DA and DB white dwarfs, where he relates the non-sinusoidal light variations with the interchanges between the convective transport of energy and the radiative one that take place in the deeper layer of the hydrogen ionization zone and where a non-linear relation for the temperature gradient is used. Regardless, in the literature these are generally referred as \textit{non-linear mixing processes} \citep{2014ApJ...783...89B} or gathered in what is called the \textit{non-linear distortion model} (\citeauthor{Bowman2016AmplitudeTargets}~\citeyear{Bowman2016AmplitudeTargets, Bowman2017};~\citeauthor{2009AA...506..471D}~\citeyear{2009AA...506..471D}).

Combination frequencies have been useful for mode identification in the asteroseismic analysis of the ZZ Ceti stars \citep[e.g.][]{2005ApJ...633.1142M}. 
However, \citet{2012MNRAS.422.1092B} demonstrated that the theory of Brickhill, extended analytically in \citet{2001MNRAS.323..248W}, is not valid for main-sequence pulsating stars, like \dss~or $\beta$ Cepheids. The reason is that geometrical effects (i.e. variation in radius and surface normal) have to be taken into account, whereas these effects are negligible in white dwarfs, because of their high density. 

Therefore, since they can be a source of confusion in mode identification for main-sequence pulsating stars, they can be considered spurious frequencies and must be properly identified and removed from the power spectra. 

Usually, these combination frequencies are identified as such, when in a prewhitening procedure the extracted frequency falls inside an \ep~range (usually the Rayleigh frequency resolution \ep~= 1/T~being T the observation interval)~around a previously calculated exact combination value \citep{Zwintz2020,Saio2018,Murphy2013,2012arXiv1210.5834P,2009AA...506..471D}. The inconsistency of this reasoning, is that when the frequency does not match exactly the value of the combination the residuals after the fit will be correlated with the so selected frequency. 

The techniques for identification and extraction of combination frequencies explained in \citet{2015MNRAS.450.3015K}, \citet{Balona2012d} and \citet{Handler2006} are more sophisticated than the aforecited, but no mathematical frame for this procedure is given. 

In this paper we present a novel self-consistent method for removing combination frequencies based on an unbiased estimation of the non-linear solution aimed at yielding residuals uncorrelated with the frequencies generating these non-linearities.

Here, we are continuing the work of \citet{1996MNRAS.281..696G} where, for the first time, the Volterra expansion is proposed as an option to model the non-linearities present in some pulsating stars.

An advantage of applying this method is that when the combination frequencies have been correctly removed from the time series, a new collection of frequencies can emerge from power spectra, which were previously hidden. In addition, it can give frequency errors much more realistic than the coarse estimation given by the Rayleigh frequency resolution.  

The paper is organised as follows: in section~\ref{sec:Teo} a theoretical basis is explained. In section~\ref{sec:Met} a detailed description of the methodology is presented. The results of applying this technique to different light curves are shown in section~\ref{sec:Res} followed by a discussion of these results in section~\ref{sec:Dis}. Finally, section~\ref{sec:Con} summarises the conclusions drawn from this study and future work.

\section{Theoretical basis: General response of a non-linear system}\label{sec:Teo}

Non-linear systems can be studied and represented by the Volterra series. In the particular case where the input to a non-linear system is a single real-valued sine wave at $\omega{_0}$ with amplitude $A_0$, the output can be expressed \cite[see p.29 in][]{PRI88} by:

\begin{equation}\label{eq:non-linear_output1}
\begin{split}
Y(t) = A_0\cdot{\Gamma_1(\omega_0)}\cdot{e^{i\cdot{\omega_0}\cdot{t}+\phi_0}}
+A_0^2\cdot{\Gamma_2(\omega_0,\omega_0)}\cdot{e^{2\cdot{i}\cdot{\omega_0}\cdot{t}+2\cdot\phi_0}}\\
+A_0^3\cdot{\Gamma_3(\omega_0,\omega_0,\omega_0)}\cdot{e^{3\cdot{i}\cdot{\omega_0}\cdot{t}+3\cdot\phi_0}}+ \dots
\end{split}
\end{equation}

Where the \Ga~functions are the generalised transfer complex functions. Sub-indexes denote the non-linear order of interactions, i.e. $\Gamma_{1}$ represents the system response for each independent frequency, $\Gamma_{2}$  represents the system response for first order interactions, and so on. 

The generalised transfer functions \Ga~can be written as Volterra expansion. These series, although they retain some of the system memory, are non-orthogonal. 

To disentangle the correlation between the basis functionals of the Volterra Series, Wiener orthogonal expansion \citep{Wiener58} can be used. The ulterior intention is to obtain a physical model that explains non-linearities. 

Current efforts are being made towards the aforementioned objective and it will be the aim of future work, but in this paper orthogonality of expression (\ref{eq:non-linear_output1}) is assumed, enabling to study certain properties that can be observed in a first order approximation. 

Although Y(t) is not linear between the input/output spectra, it is linear between Y(t) and the \Ga~ \citep{Scargle2020}, so a standard Least-Square procedure will quantify, not only the parameters of the input components, but also the contribution from the generalised transfer functions.

For example, since the \Ga~functions are complex functions, Eq.\ref{eq:non-linear_output1} can be rearranged as:
\begin{equation}\label{eq:non-linear_output2}
\begin{split}
Y(t) = \tilde{A_1}\cdot{e^{i\omega_0\cdot{t}+\tilde{\phi_{1}}}}+
       \tilde{A_2}\cdot{e^{2\cdot{i\cdot\omega_0\cdot{t}+\tilde{\phi_{2}}}}}+\\ \tilde{A_3}\cdot{e^{3\cdot{i\cdot\omega_0\cdot{t}+\tilde{\phi_{3}}}}}+...\\
\end{split}
\end{equation}

Where, 
\begin{equation}\label{eq:non-linear_output3}
\begin{split}
\tilde A_1 = A_0 \cdot |\Gamma_1(\omega_0)|,\\
\tilde A_2 = A_0^2  \cdot |\Gamma_2(\omega_0,\omega_0)|,\\
\tilde A_3 = A_0^3 \cdot  |\Gamma_3(\omega_0,\omega_0,\omega_0)|\\
\end{split}
\end{equation}
And 
\begin{equation}\label{eq:non-linear_output4}
\begin{split}
\tilde{\phi_{1}}= \phi_0+arg\lbrace\Gamma_1(\omega_0)\rbrace, \\
\tilde{\phi_{2}} = 2\cdot\phi_0+arg\lbrace\Gamma_2(\omega_0,\omega_0)\rbrace, \\
\tilde{\phi_{3}}= 3\cdot\phi_0+arg\lbrace\Gamma_3(\omega_0,\omega_0,\omega_0)\rbrace
\end{split}
\end{equation}

Examining the case when the input is composed of two real-valued sine waves at frequencies $\omega{_0}$ and $\omega{_1}$, the output of a non-linear system modeled by a Volterra series is:

\begin{equation}\label{eq:non-linear_output5}
\begin{split}
Y(t) = A_0\cdot{\Gamma_1(\omega_0)}\cdot{e^{i\cdot{\omega_0}\cdot{t}+\phi_0}}
+A_1\cdot{\Gamma_1(\omega_1)}\cdot{e^{i\cdot{\omega_1}\cdot{t}+\phi_1}}\\
+A_0^2\cdot{\Gamma_2(\omega_0,\omega_0)}\cdot{e^{2\cdot{i}\cdot{\omega_0}\cdot{t}+2\cdot\phi_0}}+A_1^2\cdot{\Gamma_2(\omega_1,\omega_1)}\cdot{e^{2\cdot{i}\cdot{\omega_1}\cdot{t}+2\cdot\phi_1}}\\
+A_0\cdot{A_1}\cdot{\Gamma_2(\omega_0,\pm{\omega_1})}\cdot{e^{i\cdot{(\omega_0\pm{\omega_1})}\cdot{t}+(\phi_0\pm{\phi_1})}}\\
+A_1\cdot{A_0}\cdot{\Gamma_2(\omega_1,\pm{\omega_0})}\cdot{e^{i\cdot{(\omega_1\pm{\omega_0})}\cdot{t}+(\phi_1\pm{\phi_0})}}+...
\end{split}
\end{equation}

Eq.\ref{eq:non-linear_output5} can be rearranged as:

\begin{equation}\label{eq:non-linear_output6}
\begin{split}
Y(t) = \tilde{A_1}\cdot{e^{i\cdot{\omega_0\cdot{t}+\tilde{\phi_{1}}}}}+
       \tilde{A_2}\cdot{e^{i\cdot{\omega_1\cdot{t}+\tilde{\phi_{2}}}}}+\\
       \tilde{A_3}\cdot{e^{2\cdot{i\cdot\omega_0\cdot{t}+\tilde{\phi_{3}}}}}+ \tilde{A_4}\cdot{e^{2\cdot{i\cdot\omega_1\cdot{t}+\tilde{\phi_{4}}}}}+\\
       \tilde{A_5}\cdot{e^{i\cdot{(\omega_0\pm{\omega_1})\cdot{t}\pm{\tilde{\phi_{5}}}}}}+...\\
\end{split}
\end{equation}

where, 
\begin{equation}\label{eq:non-linear_output7}
\begin{split}
\tilde{A_{1}} = A_{0} |\Gamma_{1}(\omega_{0})|,\\
\tilde{A_{2}} = A_{1} |\Gamma_{1}(\omega_{1}|,\\
\tilde{A_{3}} = A_{0}^{2} |\Gamma_{2}(\omega_{0},\omega_{0})|\\
\end{split}
\qquad
\begin{split}
\tilde{A_{4}} = A_{1}^{2} |\Gamma_{2}(\omega_{1},\omega_{1})|,\\
\tilde{A_{5}} =  A_{0} A_{1} |\Gamma_{2}(\omega_{0},\omega_{1})| ,\\
or~\tilde{A_{5}} =  A_{1} A_{0} |\Gamma_{2}(\omega_{1},\omega_{0})|~^{*1}
\end{split}
\end{equation}
And 
\begin{equation}\label{eq:non-linear_output8}
\begin{split}
\tilde{\phi_{1}}= \phi_{0}+arg\lbrace\Gamma_{1}(\omega_{0})\rbrace, \\
\tilde{\phi_{2}}= \phi_{1}+arg\lbrace\Gamma_{1}(\omega_{1})\rbrace, \\
\tilde{\phi_{3}}=2\cdot\phi_{0}+arg\lbrace\Gamma_{2}(\omega_{0},\omega_{0})\rbrace, \\
\end{split}
\qquad
\begin{split}
\tilde{\phi_{4}}= 2\cdot\phi_1+arg\lbrace\Gamma_2(\omega_1,\omega_1)\rbrace , \\
\tilde{\phi_{5}}= \phi_0\pm{\phi_1}+arg\lbrace\Gamma_2(\omega_0,\pm\omega_1)
\rbrace, \\ or~\tilde{\phi_{5}}= \phi_1\pm{\phi_0}+arg\lbrace\Gamma_2(\omega_1,\pm\omega_0)\rbrace^{*1}
\end{split}
\end{equation}

$^{*1}$: since no condition of symmetry is yet imposed. \newline
\newline
These amplitude and phase relations are considerably different to the ones suggested by the so called \textit{simple model} in \citet{2014ApJ...783...89B}. Analysis of amplitude and phase relations affected by the \Ga~functions (Eqs. \ref{eq:non-linear_output3},\ref{eq:non-linear_output7} and \ref{eq:non-linear_output4},\ref{eq:non-linear_output8} for the mono-periodic and double mode cases respectively) are going to be studied in the second part of this publication. 

Notice that the generalised transfer functions will not influence the frequencies, topic of this first part of the overall study. What is relevant of this formulation for the present paper is that this output could describe the rapid increases and slow decreases of the luminosity observed in some light curves.

A Fourier Transform of Eq.\ref{eq:non-linear_output6} would result in a power spectrum with peaks at the independent frequencies $\omega_0$ and $\omega_1$ (called parent frequencies in this paper), as well as on the frequencies of the non-linear terms, which are the combination frequencies (called children in this paper). This is a well-known phenomenon in system and signal analysis called \textit{intermodulation distortion} (in the case of one parent frequency is called \textit{frequency multiplication} or \textit{harmonic distortion}). 

Modelling non-linearities by treating the variable star as a non-linear but stationary system is the mathematical foundation for the method here presented, assuming a known input equal to a basic harmonic signal.  

\section{Methodology: The 'best' parents}\label{sec:Met}

Under the hypothesis that Eqs.\ref{eq:non-linear_output2} and \ref{eq:non-linear_output6} are approximation functions modelling the non-linearities present in the light curves of a mono-periodic and double-mode variable stars respectively, a least-square fit of them should yield residuals uncorrelated with the combination frequencies.

The variance of such residuals (in comparison with the variance of the original light curve) quantifies to which extend the parents and children fitted explain the signal as non-linearities: the lower the variance value the better the fit.

We define V as the continuous function of the variance related to different fittings of parent frequencies (and their statistically significant combinations) to a given light curve. Therefore, V is only a function of the parent frequencies. The aim is to find a minimum in V. 

There are several procedures for finding the minimum in a non-linear least-square fitting (when the fitted functions are non-linear in the parameters) \citep{Bevington2003}, but in the method here presented we follow an empirical approach which consists in exploring the n-dimensional independent frequencies surface.

In the case of a single parent frequency (e.g. a Cepheid or a  high amplitude mono-periodic variable), we select the highest peak in the power spectrum as a first approximation to the parent frequency with a coarse precision. Then, V($\omega$) will be sampled with a step progressively smaller until the minimum is reached. 

In this way, the minimum V value will give us a much more precise frequency. Often, the frequency error will be smaller than the nominal 1/T Rayleigh dispersion. As we will see in Section~\ref{sec:mono_Res}, the frequency value will be compatible with that obtained by the O-C method \citep{2005ASPC..335....3S}, which is basically how the times of arrival of the pulses are determined in the Pulsar Timing analysis technique, known for yielding extraordinary precision when calculating the period of a pulsar \citep{2004hpa..book.....L}.

Now that we know the 'best' parent frequency, in the case of a single parent mode, we calculate the set of combinations (in this case, harmonics) not exceeding the Nyquist frequency, to test whether they are statistically significant by giving any statistical test (e.g. Student's t or Snedecor's-F). It is important to highlight that the applicability of this method is not just for evenly-spaced data. When dealing with unequally-spaced data, the set of combinations to calculate and fit can be accordingly chosen up to any convenient frequency.  

In the case of two parent frequencies, e.g. double-mode Cepheids, High Amplitude \dss\ (HADS) or RRLyrae, or multiperiodic stars such as Low Amplitude \dss\ (LADS), \gdor, etc.) the procedure is the same for finding the 'best' parents but for calculating all the combination values we use the form:
\begin{equation}{\label{eq:combi}}
\omega_k = \left|\pm{n}\cdot{\omega_i}\pm{m}\cdot{\omega_j}\right|
\end{equation}
Where m and n are integer numbers under the condition that $\omega_k$ don't go beyond the Nyquist frequency. The absolute sum of n and m defines the combination order. We choose to limit equation (\ref{eq:combi}) to a two-termed expression to avoid increasing the probabilities for false identifications. 

A fit to the light curve of all the calculated combinations and their parents is computed. In this paper, a Student's t criterion of significance with a level of confidence of 99.9\% is used.

Finally, a Fast Fourier Transform (FFT) of the residual light curve will result in the power spectrum free of combination frequencies which is the aim of this work.

In summary, the method consists essentially in two steps: finding the parent frequencies that best describe the signal's non-linearities (we call this first step the 'best' parent method, BPM for now on), and then fitting them together with the set of children frequencies originated by them. 

\section{Results}\label{sec:Res}

Although this non-linearity analysis applies to every type of pulsator with combination frequencies, we focused in \dss\ due to  their dense and unexplained power spectra \citep{POR09,MAN12,AGH13}.

To test its performance, three \dss, of different pulsational content, were chosen:
First, a mono-periodic \ds\ variable, in order to verify the process of finding the 'best' parent. Then, a double-mode pulsator, and a multi-periodic \dss, where more complex non-linearities can be present and where their extraction could be critical to the frequency analysis. 

Details of the physical parameters and relevant information describing the light curves are listed in Table~\ref{tab:appx2_params}, located in the Appendix \ref{sec:appxB}.

It is known that gaps in time series are a source of uncertainties and error in harmonic analysis, see for instance \citet{2018AA...614A..40P}.
Therefore, gaps in the light curves used in this study where filled with MIARMA algorithm  \citep{PG15a}, which remove the effects of the observational window while  preserving the frequency content of the star. When the gap is sufficiently small (e.g. in the KIC 5950759 Kepler light curve only a few points are missing), then a linear interpolation was enough.

\subsection{Mono-periodic Stars: The \ds\ case}\label{sec:mono_Res}

\begin{figure}
	\includegraphics[width=\columnwidth]{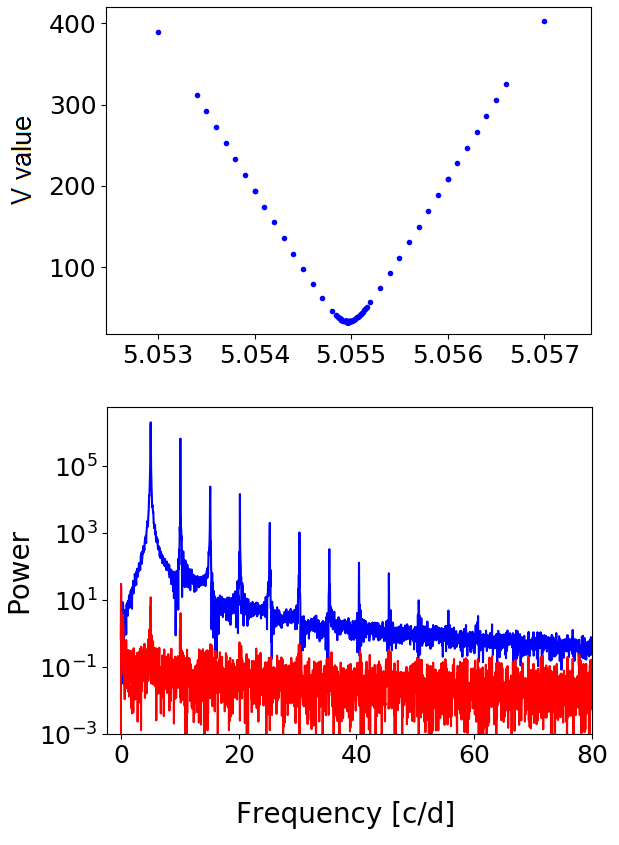}
    \caption{Upper panel: Fundamental period found as a local minimum at 5.054963644 c/d. Lower panel: in blue, the FFT of the original light curve. In Red, the FFT of the residuals after fitting the fundamental frequency and 13 statistically significant harmonics.}
    \label{fig:res_monoperiodic}
\end{figure}

The method applied to a light curve of a mono-periodic pulsating star results in a local minimum of the V function, which determines the fundamental period. This is shown for the mono-periodic \ds\ star TIC 9632550 (see Fig.~\ref{fig:res_monoperiodic}, upper panel), observed by The Transiting Exoplanet Survey Satellite \citep[TESS,][]{Ricker2014}.


\begin{table}
	\centering
	\caption{The 'best' parent search tree for the mono-periodic \ds~star TIC 9632550. First column quantifies the number of statistically significant frequencies, or children, detected with the parent frequency specified in the third column, in cycles per days units (zeros omitted for the sake of clarity). Second column is the variance after fitting the parent and combination frequencies (in this case, only harmonics of the highest one).}
	\label{tab:res1_table1}
	\begin{tabular}{ccc} 
		\hline
		N of fitted &V value & Frequency[c/d]\\\raisebox{1.0ex}{ frequencies}&&\\
		\hline
        1  &3155.844405793707201 &5.0 \\
        5  &948.937738175725485 &5.05 \\
        14 &33.629889361388315 &5.055 \\
        14 &33.629889361388315 &5.055\\
        \textbf{14}&\textbf{32.871889584474623} &\textbf{5.05496}\\
        14 &32.862102488776905 &5.054964\\
        14 &32.862102488776905 &5.054964\\
        14 &32.862101660828912 &5.05496404\\
        14 &32.862101656257245 &5.054964037\\
        14 &32.862101656225789 &5.0549640372\\
        14 &32.862101656224368 &5.05496403722\\
        14 &32.862101656224311 &5.054964037229\\
        14 &32.862101656223409 &5.0549640372274\\
        14 &32.862101656222627 &5.05496403722726\\
        14 &32.862101656222627 &5.05496403722726\\
		\hline
	\end{tabular}
\end{table}

\begin{table}
	\centering
	\caption{Results of the combination frequencies extraction process for the mono-periodic \ds~star TIC 9632550. First column show the 'best' parent from the search tree in cycles per day. Second column specifies the number of statistically significant frequencies, or children, extracted. The \%CF (third column) quantifies the percentage of initial power due to the combination frequencies and their parents.}
	\label{tab:res1_table2}
	\begin{tabular}{cccc} 
		\hline
		\multicolumn{4}{c}{TIC 9632550} \\
		\hline
		 Tag &'best' parent [c/d] & Combinations & \%CF \\ &&\raisebox{1.0ex}{extracted}&\\
		\hline
         f0 & 5.05496 & 13 Harmonics & 98.98  \\
		\hline
	\end{tabular}
\end{table}

The iterative process of searching for the 'best' parent explained in Section \ref{sec:Met}, refines the frequency until the variance value (V) does not change. In this particular case, it happened in the 14$^{th}$ iteration (see Table~\ref{tab:res1_table1}). However, the list is truncated at the 5$^{th}$ iteration due to the numerical uncertainty (see Appendix ~\ref{sec:appxA} for a full explanation).

\begin{figure}
	\includegraphics[width=\columnwidth]{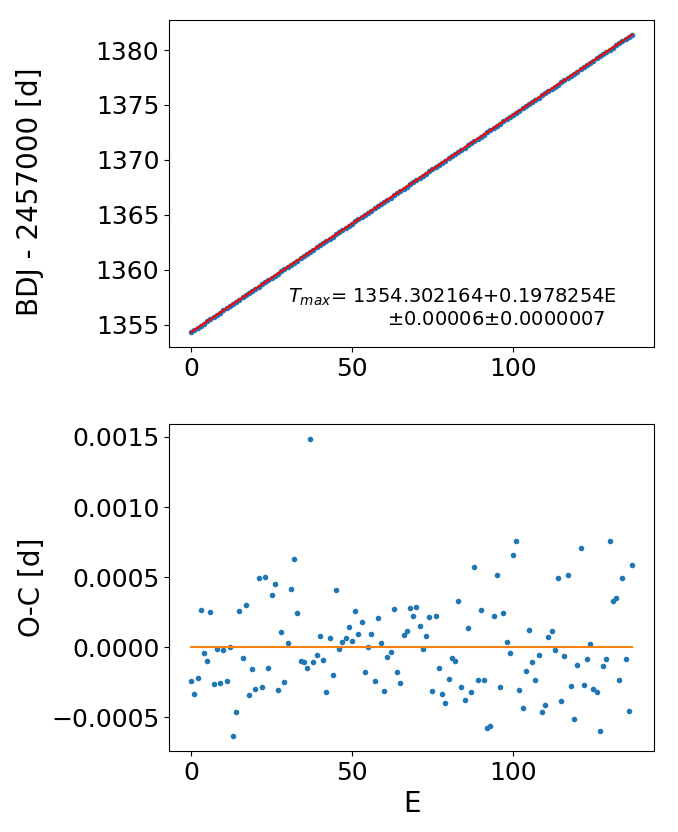}
    \caption{Upper panel: in blue, the times of the light maximum, corresponding to the maximum value of a parabola fitted to each cycle.  Red: Regression line $T_{max} = T_0+PE$, where P is the trial period, $T_0$ is the zero epoch, and E an integer number of cycles elapsed since the zero epoch. Fundamental frequency ($\omega_0=1/P$) calculated by the O-C method: $\omega_0 = (5.05496 \pm 0.00002)~c/d.$ Lower panel: residuals of the regression.}
    \label{fig:O-C}
\end{figure}

Our result for the fundamental frequency is compatible with the one obtained by the O-C procedure (see Figure~\ref{fig:O-C}):\\  \centerline{$\omega_0$ = (5.05496 $\pm$ 0.00002) c/d.}

The FFT of the residual light curve after fitting the fundamental frequency and its harmonics is almost cleaned from all signal contribution. This is graphically represented in the lower panel of Figure~\ref{fig:res_monoperiodic} and quantitatively expressed by the high percentage of the original power explained by these non-linear effects and their parents (expressed as \%CF in Table~\ref{tab:res1_table2}).

There are three peaks remaining in the residual power spectrum: the first one correspond to the first frequency bin and it is the residual of a second order polynomial fitting performed in order to remove any trend in the light curve; the second peak, appearing next to the fundamental frequency, and the third one, which is next to the first harmonic, are possibly explained by an amplitude modulation of the fundamental frequency. An alternative explanation of these two peaks appearing at about the same frequencies as the fitted ones, might be the fractal property studied by \citet{DeFranciscis2018}, which is impossible to reproduce by using a Fourier representation. In any case, the logarithmic scale shows 5 orders of magnitude difference between the original and the residual power, which is in very good agreement with our expectation of uncorrelated residuals.

\subsection{Double mode stars: The HADS case}
We will now extend the procedure described in the previous section for the 'best' parent search to double mode stars.

There is a slight tendency for HADS stars to have higher number of non-linearities \citep{2016MNRAS.459.1097B}. This is why we chose to apply the method to KIC 5950759, a HADS star observed by the Kepler satellite \citep{Gilliland2010}, whose non-linearities were first studied by \citet{Bowman2017}.


\begin{figure}
	\includegraphics[width=\columnwidth]{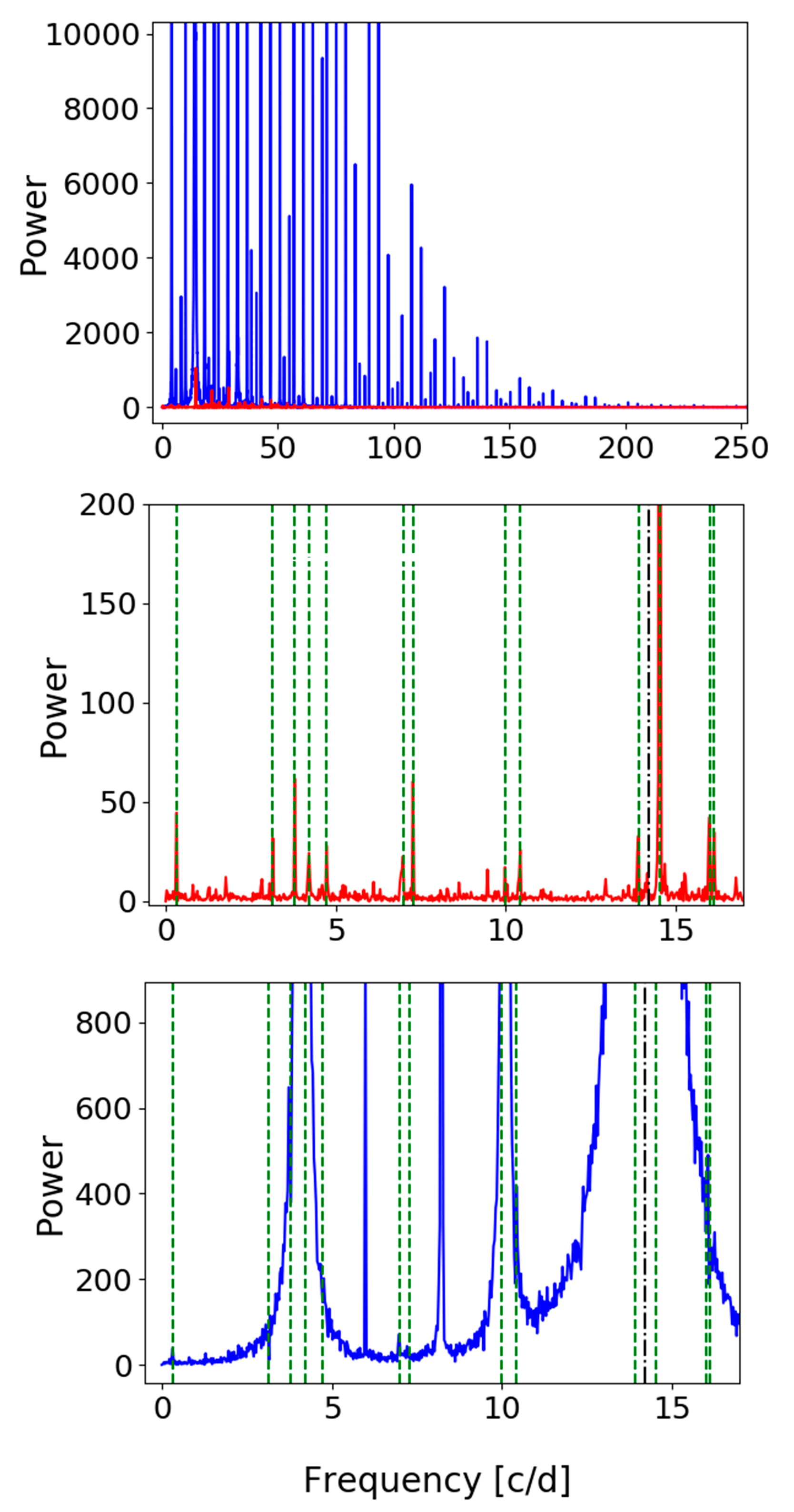}
    \caption{Blue: FFT of the original light curve. Red: FFT of the residuals after fitting  the 'best' parents and the combinations generated by them. Green dashed: new frequencies detected in the residuals of the fitting. Black dash-dotted: 'best' parent frequencies. Notice in middle panel the significant peak corresponding to $\omega_m\approx0.32$ c/d}
    \label{fig:res_doublemode}
\end{figure}

The original power spectrum (blue in Figure~\ref{fig:res_doublemode}) shows a very regular structure where the first two highest peaks follow the fundamental period and first overtone ratio expected to occur for \dss\ \citep[see][]{1979ApJ...227..935S}.

\begin{table}
	\centering
	\caption{Results of the combination frequencies extraction process for the double-mode HADS star KIC 5950759. First column show the 'best' parents from the search tree in cycles per day. Second column specifies the number of statistically significant frequencies, or children, extracted. The \%CF (third column) quantifies the percentage of initial power due to the combination frequencies and their parents.}
	\label{tab:res2_table1}
	\begin{tabular}{cccc} 
		\hline
		\multicolumn{4}{c}{KIC 5950759} \\
		\hline
		 Tag &'best' parents [c/d] & Combinations & \%CF \\ &&\raisebox{1.0ex}{extracted}&\\
		\hline
         f0&14.22136&177 in total:&97.48 \\
         \raisebox{1.0ex}{f1}&\raisebox{1.0ex}{18.33722}&\raisebox{1.0ex}{17 harmonics}&\\
         &&\raisebox{1.0ex}{92 sums}&\\
         &&\raisebox{1.0ex}{and 68 differences}&\\
		\hline
	\end{tabular}
\end{table}

The 'best' parents (see Table \ref{tab:res2_table1}) are compatible with the ones presented by \citet{2018ApJ...863..195Y} for the SC data:\\ \centerline{$\omega_0=(14.221367\pm{0.000015})~c/d$} \\ \centerline{$\omega_1=(18.337228 \pm{0.000023})~c/d $}

These frequency errors were calculated according to a heuristically derived formula for the upper limit of the frequency uncertainty in \citet{2008AA...481..571K}. The precision reached with the 'best' parents search is also empirically justified in Appendix~\ref{sec:appxA}.

Due to the high power of the initial components, the fitting was completed in three steps, continuing the extraction in the residual light curve. After the fitting, almost every contribution from combination frequencies where effectively extracted and, as a consequence, a new frequency structure (that was previously hidden) emerges in its power spectrum (see middle panel of  Fig.~\ref{fig:res_doublemode}).

The frequency  $\omega_m\approx0.32$ c/d , first claimed by \citet{2018ApJ...863..195Y} to be modulating the entire spectrum, is now significant according to the \citet{Reegen07} criteria of signal/noise > 12.57 in the power domain. 

As firstly discussed by \citet{Bowman2017}, the new frequency structure seen for the KIC 5950759 deserves further studies regarding its origin since it could be possibly indicating the the existence of other independent modes. \citet{2018ApJ...863..195Y} explored several physical explanations, proposing as the most likely to identify  $\omega_m$ as the rotation of the star, meaning that the dashed lines around $\omega_0$ in the middle panel of Figure~\ref{fig:res_doublemode} corresponds to the modulation of the main pulsation modes with rotation frequency. But the high power for the frequency  $\omega_0$+$\omega_m$ suggests that it could still be an independent mode, resonantly coupled with the combination coming from the modulation effect. This possibility is going to be analysed in upcoming studies.
 
In any case, the results shown in  Figure~\ref{fig:res_doublemode} provide a new level of relevance to this method: with a correct extraction of the combination frequencies, like the one proposed in this work, it is possible to unveil frequency structures that previously did not exceed the detection threshold. In this particular case, we were able to detect the $\omega_m$ frequency using short cadence (SC) data, while in \citet{2018ApJ...863..195Y} long cadence (LC) data, a super-Nyquist and alias analysis were necessary in order to identify it.

\subsection{Multi-periodic stars: The LADS case.}

For more than two parents, exploring the frequency space of the V function recursively to find its minimum value can be computationally expensive, but the implications of applying the BPM can be crucial for an asteroseismic analysis, as we showed in the previous section. 

\begin{figure}
	\includegraphics[width=\columnwidth]{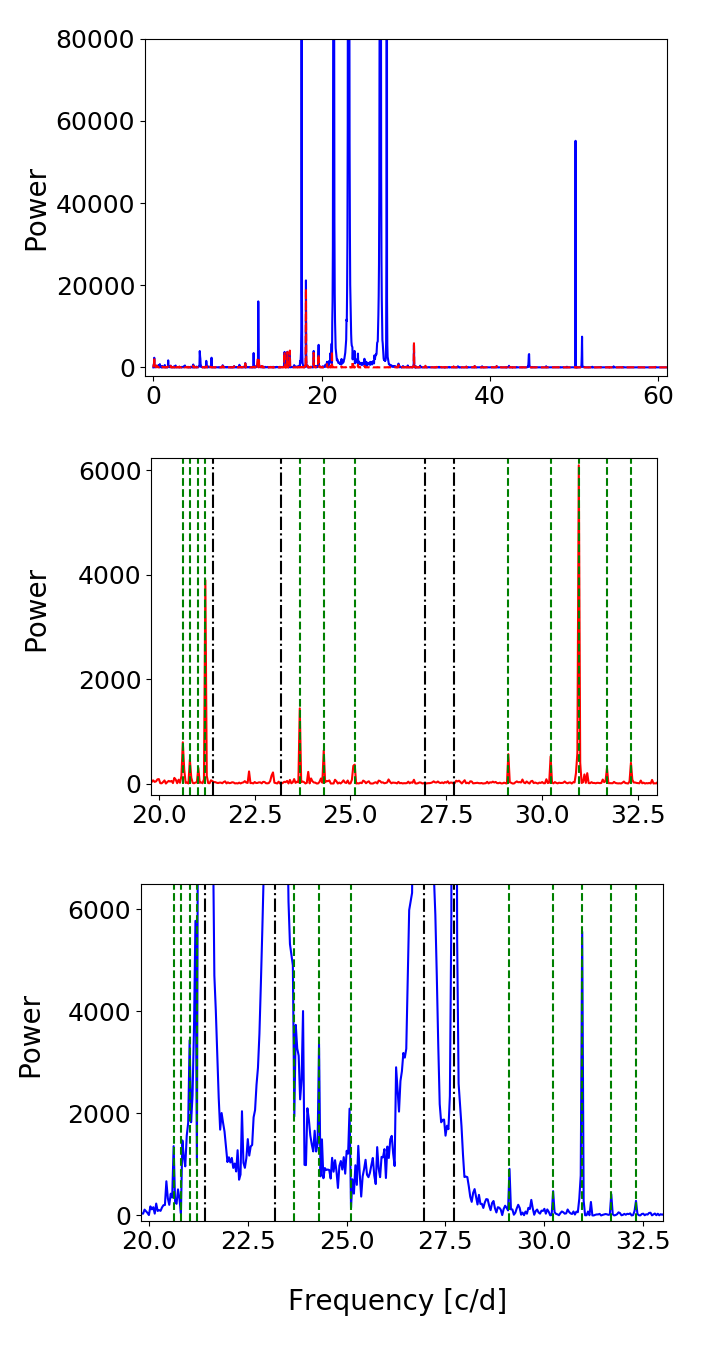}
    \caption{Blue: FFT of the original light curve. Red: FFT of the residuals after fitting  the 'best' parents and  the series of combinations originated by them. Green dashed: new frequencies detected after the fit. Black dash-dotted: 'best' parent frequencies.}
    \label{fig:res_multiperiodic}
\end{figure}

We applied the method to the light curve of HD 174966 (see Fig.~\ref{fig:res_multiperiodic}), a LADS (or simply \ds~star) observed by the CoRoT satellite \citep{Auvergne2009d}. This star was studied in \citet{AGH13}, who found that the 5th highest peak in the amplitude spectrum was very near to the estimated fundamental radial mode ($17.3\pm 2.5$~c/d).

Therefore, we choose as independent frequencies the first five peaks of highest power but, instead of searching the combinations of the full set of five independent frequencies, we decided to divide the search in couples in order to reduce the computational cost of the BPM solution. 

\begin{table}
	\centering
	\caption{BPM for every possible couple of the first five high power peaks in the HD 174966 power spectrum}
	\label{tab:res3_table0}
	\begin{tabular}{cc} 
		\hline
		Couple Tag & Couple Frequencies [c/d] \\
		\hline
		(f0, f1) & (17.62288, 23.19479) \\
		(f0, f2) & (17.62298, 21.42079) \\
		(f0, f3) & (17.62280, 26.95853) \\
		(f0, f4) & (17.62291, 27.71456) \\
		(f1, f2) & (23.19477, 21.42097) \\
		(f1, f3) & (23.19479, 26.95851) \\
		(f1, f4) & (23.19477, 27.71503) \\
		(f2, f3) & (21.42078, 26.95853) \\
		(f2, f4) & (21.42078, 27.71463) \\
		(f3, f4) & (26.95851 , 27.71487) \\
		\hline
	\end{tabular}
\end{table}

In Table \ref{tab:res3_table0}, the results of the BPM for every possible couple with the precision adopted in this work (see \ref{sec:appxA}) show small differences. In this case, we select as the 'best' parents for the 5 highest amplitude peaks the values with the best precision in which they are compatible in each search.


\begin{table}
	\centering
	\caption{Results of the combination frequencies extraction process for the multi-mode \ds~star HD 174966. First column show the 'best' parents from the search tree in cycles per day. Second column specifies the number of statistically significant frequencies, or children, extracted. The \%CF (third column) quantifies the percentage of initial power due to the combination frequencies and their parents.}
	\label{tab:res3_table1}
	\begin{tabular}{cccc} 
		\hline
		\multicolumn{4}{c}{HD 174966} \\
		\hline
		Tag &'best' parents [c/d] & Combinations & \%CF \\ &&\raisebox{1.0ex}{extracted}&\\
		\hline
         f0 & 17.6230 &118 in total:&93.01 \\
         \raisebox{1.0ex}{f1}&\raisebox{1.0ex}{23.1948}&\raisebox{1.0ex}{1 harmonic}&\\
        \raisebox{1.0ex}{f2} &\raisebox{1.0ex}{21.4210}&\raisebox{1.0ex}{11 sums}&\\
        \raisebox{1.0ex}{f3} &\raisebox{1.0ex}{26.9585}&\raisebox{1.0ex}{106 differences}&\\
         \raisebox{1.0ex}{f4} &\raisebox{1.0ex}{27.7150}&&\\
		\hline
	\end{tabular}
\end{table}

Particularly in this star, 118 combinations were statistically significant (see Table \ref{tab:res3_table1}). Notice that the number of differences is higher than the number of sum combinations. This may raise concerns about the reality of these identifications in the lower frequency range.

The majority of these differences correspond to high order combinations (see Table \ref{tab:appx3_table3}), but harmonics ($n\cdot\omega_i$ or $m\cdot\omega_j$) with such high n and m, are not statistically significant (only 2f3 is detected). Alternatively, these significant differences could simply be false identifications due to the fact that at higher n and m more combinations are tested, increasing the probability of a match, as well as the possibility of choosing as parent frequencies a combination frequency. 

In this work, we do not exclude any match since we are interested in finding the set of combination frequencies that could explain the most of the signal as non-linearities. Besides, \citet{2015MNRAS.450.3015K} states that the amplitude of a child frequency could be higher than their parents amplitudes, which could explain the missing high order harmonics issue. 

As previously mentioned, further discriminating criteria (apart from their frequency value) are required for an unambiguous identification, in this way avoiding false identifications due to high values of n and m. 

Nevertheless, this example shows that in no case the method is introducing new frequencies and that even when is not clear that the arbitrarily chosen as parent frequencies are actual oscillation modes of the star (which is often the case when dealing with multi-periodic star), the set of significant combinations resulting from the algorithm can still be useful to test if extracting them has simplified the power spectrum in agreement with a solution from a linear stellar oscillation model (see Fig. \ref{fig:res_multiperiodic}, where some of the green dashed lines are equally spaced, possibly identifiable with non-radial frequency structures or rotational splittings).



\section{Discussion : Frequencies inside the Rayleigh interval and Biased Least-Square solutions.}\label{sec:Dis}



In this paper, we focus in discriminating combination frequencies from oscillation modes of the star only by the frequency relationship between parents and children (see Eq.\ref{eq:combi}). However, pulsation modes could still have the same frequency value as a combination by chance, being this issue the main limitation of this method. Other considerations regarding the phases and amplitudes of the non-linearities, described by the generalised transfer functions \Ga,  will be examined in upcoming studies, looking for an unambiguous identification of a non-linearity.

Nevertheless, this study revealed the necessity of revising some common misconceptions in the field in order to obtain well-defined results.

First of all, we consider spurious peaks as significant and non-significant maxima in the frequency spectrum that do not correspond to any oscillation mode of the pulsating star\citep{Suarez2020}. In this sense, peaks associated to non-linear effects should not be considered spurious since they are originated by oscillations themselves. However, no closed-form formula like, for instance, Eqs.~\ref{eq:non-linear_output1} and \ref{eq:non-linear_output5}, have been obtained so far to characterise these non-linear effects. Therefore, until such expression is developed, we can consider non-linear components appearing in the frequency spectrum as spurious peaks.

In practice, the Rayleigh frequency resolution is often used as the error for the identification of combination frequencies. Frequencies extracted from a prewhitening cascade are found to be a combination when they fall inside the Rayleigh frequency range. However, after some numerical exercise performed for this particular case, we have found that fitting different sets of combination frequencies inside the Rayleigh interval, showed significantly different residuals. Therefore, Rayleigh frequency dispersion cannot be considered as a real frequency error.

The Rayleigh frequency resolution becomes a good uncertainty estimator when dealing with frequencies closer than $1/T$ to each other, but it could be exceeded when there is only one frequency component inside the Rayleigh interval. Resolution is not the same as precision.

On the other hand, extracting combination frequencies in a least-square sense as a first step before undertaking the frequency analysis can expose  pulsation modes or frequency spacing patterns in which we are interested on, otherwise hidden. We have verified this in the results of the double-mode HADS, where the modulating frequency $\omega_m$ could be detected.


Finally, it is important to note that minimising the residuals does not guarantee that a real solution has been found. A least-square fit, exploring all free parameters (frequencies, amplitudes and phases) without preserving that the combinations are integer times the parent frequencies, can result in smaller residuals, but these cannot be explained with a closed-form formula.

\section{Conclusions and future work}\label{sec:Con}

Fundamental frequencies of mono-periodic stars analysed by this method are compatible with those calculated with the commonly used O-C method. The extension to stars pulsating in two or more parent frequencies is envisaged in this manuscript.

Although we support a conservative approach regarding the numerical precision (of the order of $\approx10^{-5} d^{-1}$, see Sec.\ref{sec:appxA}), the method yields more realistic frequency uncertainties than the usual Rayleigh dispersion, a rough estimator of the frequency error.

Moreover, fitting combination frequencies with this method could help to identify pulsation modes by possibly revealing radial and non-radial frequency patterns or rotational splittings in the periodogram of the residuals. An example of this was shown in the results for the SC observations of the HADS star KIC 5950759, where the modulating frequency $\omega_{m}$ found by \citet{2018ApJ...863..195Y} became detectable.

In the light of these conclusions, some common misconceptions when analysing non-linearities should be reconsidered: the Rayleigh dispersion is not necessarily the inaccuracy associated to the determination of combination frequencies $\omega_{k}$ (\ref{eq:combi}).

This method partially overcomes the necessity to correct non-linearities in order to achieve an efficient mode identification with a linear pulsation model. Future research could explore the link between the theoretical foundations of this paper and the new MESA functionality regarding non-linear radial stellar pulsations \citep{Paxton2019}. 



In essence, we can now identify the set of combination frequencies that best describe the signal, but the method do not supply a physical mechanism to explain their visibilities. There is still a possibility for the combination frequency to not correspond to an interaction between modes due to their intrinsic non-linear behaviour. It could be a new unstable mode (very near the exact combination frequency value) that undergoes amplitude enhancement due to the resonance mode coupling mechanism \citep{Dzie85, Dzie88, Hoolst94}. This is an important aspect to be determined since it could clarify why stars with similar parameters exhibit very different power spectra. \citep{2011MNRAS.417..591B}. Consequently, extra criteria for an unambiguous identification constitute the aim of the second part of this paper. 

This method is shown to have a very high potential in exploring combination frequencies as the output of a non-linear system. The underlying physical meaning of the generalised transfer functions \Ga~ defined in Eqs. (\ref{eq:non-linear_output1}) and (\ref{eq:non-linear_output5}) is the next step of this research. Future studies will consider orthogonal expansion in terms of a Wiener series opening a new insight on the relationships between amplitudes and phases as first envisioned by \citet{1996MNRAS.281..696G}. The final objective is to properly characterise the non-linear response of a pulsating star.

\section*{Acknowledgements}

MLM, RGH and JPG acknowledge financial support from the State Agency for Research of the Spanish MCIU through the "Center of Excellence Severo Ochoa" award to the Instituto de Astrof\'isica de Andaluc\'ia(SEV-2017-0709). Authors acknowledge funding support from Spanish public funds for research under project ESP2017-87676-C5-5-R.




\bibliographystyle{mnras}
\bibliography{references.bib} 





\section{Appendix}
\subsection{Uncertainties in frequencies.}\label{sec:appxA}

The method described in this manuscript for studying combination frequencies mainly relies in how we determine the 'best' parents. Progressively increasing the precision in frequency (when searching for the minimum of the variance of the residuals) involves getting closer to the floating point number precision, which implies that numerical errors are an important source of uncertainty.
 
Finding when this numerical effects are hampering the 'best' parents computations, will provide us with an estimate of the upper limit in the uncertainty of the frequencies.

We find this limit by building a synthetic light curve in this way:
\begin{equation}
    S(t) = \sum_{k=1}^n A_k \cos (2\pi k\omega t + \phi_k), 
    \quad   k, n\in\mathbb{N}
\end{equation}
where $\omega$ is the parent frequency for a mono-periodic variable up to n harmonics. The input Fourier parameters of the synthetic light curve have been obtained applying the method initially to real data (the 'best' parent frequencies and their combinations). The synthetic light curve will have the same number of data points as the observations and no added noise.

The output of the method applied to the simulated light curve will have to converge to the 'best' parents initial values (i.e. the input parent and combination frequencies of the synthetic light curve). The variance at that point (V value), theoretically expected to be zero, will reveal the error in machine calculations.

\begin{table}
	\centering
	\caption{The 'best' parent search tree for the synthetic light curve build from TIC 9632550 data. First column quantifies the number of statistically significant frequencies, or children, detected with the parent frequency specified in the third column, in cycles per days units (zeros omitted for the sake of clarity). Second column is the variance after the fit of the parent and combination frequencies (in this case, only harmonics of the highest one).}
	\label{tab:appx1_table1}
	\begin{tabular}{ccc}
		\hline
		N of fitted &V value & Frequency [c/d]\\\raisebox{1.0ex}{ frequencies}&&\\
		\hline
		1&3156.591456884018044 & 5.0 \\
		5& 948.685924387723073 & 5.05 \\
		14&  7.165213986246192 & 5.055 \\
		14&  7.165213986246192 & 5.055 \\
		14&  0.804372110830800 & 5.05496 \\		
		14&  0.007417117763935 & 5.054964\\
		14&  0.007417117763935  & 5.054964\\
		14&  0.000552438323944  & 5.05496404 \\
		14&  0.000045278362580 & 5.054964037 \\
		14&  0.000005430569507 & 5.0549640372 \\
		14&  0.000000546520136 & 5.05496403723 \\
		14&  0.000000051236668 & 5.054964037227 \\
		14&  0.000000008582531 & 5.0549640372273 \\
		\textbf{14}&\textbf{6.02951e-10} & 	\textbf{5.05496403722726 }\\
		\hline
	\end{tabular}
\end{table}

Results of this test using the components extracted form TIC 9632550, the mono-periodic \ds~star observed by TESS, are listed in Table \ref{tab:appx1_table1}. The initial 'best' parent is reached with the V value of order $\approx6\cdot10^{-10}$. Consequently, V values smaller than this number are compromised by the numerical errors intrinsic to the machine calculations.

\begin{table}
	\centering
	\caption{Results of the fundamental frequency determination by the 'best' parent search and O-C method for the 4 partitions of the light curve of the mono-periodic \ds\ star, TIC 9632550. Each section is $\approx$7 days long.}
	\label{tab:appx1_table2}
	\begin{tabular}{ccc} 
		\hline
		Section & 'Best' parent [c/d] & O-C Frequency [c/d]  \\
		\hline
		S1 & 5.05491643 & $5.0548\pm 0.0001$   \\
		S2 & 5.05490929 & $5.0550\pm 0.0001$   \\
		S3 & 5.05501342 & $5.0549\pm 0.0002$ \\
		S4 & 5.05501167 & $5.0548\pm 0.0002$  \\

		\hline
	\end{tabular}
\end{table}


In addition, we divided the real light curve of TIC 9632550 in four sections and find the 'best' parent in each of this partitions. In spite of the reduced frequency resolution, due to the smaller time interval of the light curve, the parent frequency found for each partition is similar, and also compatible with the O-C method up to the 4$^{th}$ decimal (see Table \ref{tab:appx1_table2}). This test confirm the robustness of the BPM search .

In order to test if the effect of leakage had something to do with the small variation in the 4$^{th}$ decimal (i.e. $\approx$4e-6 s period variations), we applied the BPM to the mono-periodic light curve with exactly an integer number of cycles and with an integer number of cycles plus half a cycle. 

Results of this test highlighted the influence of the number of cycles on the determination of the period. Future work could explore this issue to give a precise lower limit to the frequency uncertainty. In this regard, a conservative approach is adopted in this paper expressing results with the precision that the O-C method achieves.
 

As a last remark, the duration of the observation is a relevant parameter for estimating the frequency uncertainty when dealing with two close frequencies (closer than $1/T$ to each other), and so, the Rayleigh frequency resolution becomes a good estimator. But, precision can go further the Rayleigh frequency resolution if there is only one frequency component inside the Rayleigh interval. We just proved this for each $\approx$7 days long sections of the light curve of the mono-periodic \ds star, TIC 9632550. The numerical precision reached is $\pm1\cdot10^{-8}$ and is not "as if we were observing $\approx$274000 years". Rayleigh resolution remains the same: $\approx$1/7 days, that is, $\approx$0.14 c/d. 

\subsection{Stellar parameters and relevant information of the time series used.}\label{sec:appxB}

\begin{table*}
    \centering
    \caption{Stellar parameters from Gaia DR2 catalogue and time series information from each space satellite. T is the length of the observation in days and $\delta_t$~ is the cadence or sampling rate in seconds. For the TESS and Kepler light curves, we used the instrumental effects free light curve, resulting from the Pre-Search data Conditioning (PDC) pipeline, accessible in  the  Mikulski  Archive for Space Telescopes (MAST: \url{https://archive.stsci.edu/}).}
    \label{tab:appx2_params}
	\begin{tabular}{lccc|ccc} 
    \hline
	\multicolumn{4}{c|}{Stellar}  &\multicolumn{3}{c}{Time series}\\
	\multicolumn{4}{c|}{Parameters}&\multicolumn{3}{c}{Parameters}\\
		\hline
		Name & Spectral & Magnitude & Effective Temp. & T [d] & $\delta_t$~[s] & Obs.\\
		     & Type & $m_v$ ~[mag] & $T_{\text{eff}}$~[K] &&&Sequence\\
		\hline
		&&&&&& \\
		 TIC 9632550 & A8III & $9.317 \pm 0.009$ & $ 7009.25_{-6800.90}^{+7210.64}$ & 27.41 & 120.01 & Sector 2 \\
		 &&&&&& \\
		 KIC 5950759 & -- & $13.8271 \pm{0.0141}$ & $ 7842.5_{-7595.00}^{+7995.00}$ & 31.04 & 58.85 & Quarter 4\\
		 &&&&&& \\
		 HD 174966 & A7III/IV & $ 7.6498 \pm {0.0005}$ & $ 7446.67 _{-7291.50}^{+7583.33}$  & 27.20 & 31.99 & Run SRc01 \\
		 &&&&&& \\
		\hline
		\end{tabular}
\end{table*}

\subsection{Combination frequencies detected.}\label{sec:appxc}

\begin{table}
	\centering
	\caption{Tags of the statistically significant combination  frequencies for the mono-periodic \ds~star TIC 9632550. The frequency values can be calculated with the given parent frequencies resulting from BPM since the fitted values are the exact combination frequency values}
	\label{tab:appx3_table1}
	\begin{tabular}{ccc} 
		\hline
		\multicolumn{3}{c}{Non-linearities}\\\multicolumn{3}{c}{of TIC 9362550 }\\
		\hline
         2f0 & 7f0  & 12f0 \\
         3f0 & 8f0  & 13f0 \\
         4f0 & 9f0  & 14f0 \\
         5f0 & 10f0 & \\
         6f0 & 11f0 & \\
		\hline
	\end{tabular}
\end{table}

\begin{table}
	\centering
	\caption{Tags of the statistically significant combination  frequencies for the double mode HADS star KIC 5059759. The frequency values can be calculated with the given parent frequencies resulting from BPM since the fitted values are the exact combination frequency values}
	\label{tab:appx3_table2}
	\begin{tabular}{cccccc} 
		\hline
		\multicolumn{6}{c}{Non-linearities of KIC 5059759} \\
		\hline
         2f0 & 3f0+2f1 & 7f0+5f1 & 11f0+7f1 & 6f0-3f1 & 4f1-1f0\\
         3f0 & 3f0+3f1 & 7f0+6f1 & 12f0+1f1 & 7f0-1f1 & 4f1-2f0\\
         4f0 & 3f0+4f1 & 7f0+7f1 & 12f0+2f1 & 7f0-2f1 & 4f1-3f0\\
         5f0 & 3f0+5f1 & 8f0+1f1 & 12f0+3f1 & 7f0-3f1 & 4f1-5f0\\
         6f0 & 3f0+6f1 & 8f0+2f1 & 12f0+4f1 & 7f0-4f1 & 5f1-1f0\\
         7f0 & 3f0+7f1 & 8f0+3f1 & 12f0+5f1 & 7f0-5f1 & 5f1-2f0\\
         8f0 & 4f0+1f1 & 8f0+4f1 & 12f0+6f1 & 8f0-1f1 & 5f1-3f0\\
         9f0 & 4f0+2f1 & 8f0+5f1 & 13f0+1f1 & 8f0-2f1 & 5f1-4f0\\
         10f0 & 4f0+3f1 & 8f0+6f1 & 13f0+2f1 & 8f0-4f1 & 5f1-6f0\\
         11f0 & 4f0+4f1 & 8f0+7f1 & 13f0+3f1 & 8f0-6f1 & 6f1-4f0\\
         12f0 & 4f0+5f1 & 9f0+1f1 & 13f0+4f1 & 9f0-1f1 & 6f1-5f0\\
         13f0 & 4f0+6f1 & 9f0+2f1 & 13f0+5f1 & 9f0-2f1 & 6f1-6f0\\
         14f0 & 4f0+7f1 & 9f0+3f1 & 13f0+6f1 & 9f0-6f1 & 6f1-7f0\\
         2f1  & 5f0+1f1 & 9f0+4f1 & 14f0+1f1 & 10f0-1f1 & 7f1-6f0\\
         3f1  & 5f0+2f1 & 9f0+5f1 & 14f0+2f1 & 10f0-2f1 & 7f1-7f0\\
         4f1  & 5f0+3f1 & 9f0+6f1 & 14f0+3f1 & 10f0-4f1 & 7f1-8f0\\
         5f1  & 5f0+4f1 & 9f0+7f1 & 14f0+4f1 & 10f0-6f1 & 7f1-9f0\\
         1f0+1f1 & 5f0+5f1 & 10f0+1f1 & 14f0+5f1 & 10f0-7f1 & 8f1-7f0\\
         1f0+2f1 & 5f0+6f1 & 10f0+2f1 & 15f0+4f1 & 11f0-1f1 & 8f1-6f0\\
         1f0+3f1 & 6f0+1f1 & 10f0+3f1 & 2f0-1f1 & 11f0-2f1 & 8f1-8f0\\
         1f0+4f1 & 6f0+2f1 & 10f0+4f1 & 3f0-1f1 & 11f0-7f1 & 8f1-9f0\\
         1f0+5f1 & 6f0+3f1 & 10f0+5f1 & 3f0-2f1 & 11f0-4f1 & 9f1-7f0\\
         1f0+6f1 & 6f0+4f1 & 10f0+6f1 & 4f0-1f1 & 12f0-1f1 & 9f1-8f0\\
         2f0+1f1 & 6f0+5f1 & 10f0+7f1 & 4f0-2f1 & 20f0-14f1 & 15f1-19f0\\
         2f0+2f1 & 6f0+6f1 & 11f0+1f1 & 4f0-3f1 & 1f1-1f0 & 16f1-19f0\\
         2f0+3f1 & 6f0+7f1 & 11f0+2f1 & 5f0-1f1 & 2f1-1f0 &17f1-19f0\\
         2f0+4f1 & 7f0+1f1 & 11f0+3f1 & 5f0-2f1  & 2f1-2f0 & 18f1-20f0\\
         2f0+5f1 & 7f0+2f1 & 11f0+4f1 & 5f0-3f1 & 3f1-1f0 &\\
         2f0+6f1 & 7f0+3f1 & 11f0+5f1 & 6f0-1f1 & 3f1-2f0 &\\
         3f0+1f1 & 7f0+4f1 & 11f0+6f1 & 6f0-2f1 & 3f1-3f0 &\\
		\hline
	\end{tabular}
\end{table}

\begin{table}
	\centering
	\caption{Tags of the statistically significant combination  frequencies for the multi-periodic \ds~star HD 174966. The frequency values can be calculated with the given parent frequencies resulting from BPM since the fitted values are the exact combination frequency values}
	\label{tab:appx3_table3}
	\begin{tabular}{cccc} 
		\hline
		\multicolumn{4}{c}{Non-linearities of HD 174966} \\
		\hline
         2f3     & 6f0-5f2 & 9f1-9f2 & 8f2-7f3\\
         1f0+1f1 & 7f0-6f2 & 1f1-1f3 & 9f2-6f3\\
         1f0+1f2 & 8f0-6f2 & 3f1-2f3 & 9f2-7f3\\
         1f0+1f3 & 9f0-6f2 & 3f1-5f3 & 1f2-1f4\\
         1f1+1f2 & 9f0-7f2 & 4f1-3f3 & 3f2-1f4\\
         1f1+1f3 & 9f0-8f2 & 6f1-4f3 & 4f2-2f4\\
         1f1+2f3 & 2f0-1f3 & 7f1-6f3 & 4f2-3f4\\
         1f1+1f4 & 3f0-2f3 & 9f1-8f3 & 4f2-4f4\\
         2f1+1f3 & 5f0-2f3 & 3f1-3f4 & 5f2-4f4\\
         2f1+1f4 & 5f0-4f3 & 4f1-2f4 & 6f2-3f4\\
         1f2+1f4 & 5f0-5f3 & 4f1-4f4 & 6f2-4f4\\
         1f3+1f4 & 7f0-4f3 & 5f1-4f4 & 7f2-4f4\\
         1f0-1f1 & 7f0-5f3 & 6f1-5f4 & 7f2-6f4\\
         2f0-1f1 & 8f0-5f3 & 7f1-6f4 & 8f2-6f4\\
         3f0-1f1 & 8f0-6f3 & 8f1-6f4 & 8f2-8f4\\
         3f0-3f1 & 2f0-2f4 & 8f1-7f4 & 8f2-9f4\\
         5f0-4f1 & 2f0-3f4 & 8f1-8f4 & 9f2-7f4\\
         6f0-3f1 & 3f0-1f4 & 1f2-1f3 & 9f2-8f4\\
         6f0-5f1 & 4f0-3f4 & 2f2-1f3 & 1f3-1f4\\
         7f0-6f1 & 4f0-4f4 & 3f2-2f3 & 1f3-2f4\\
         8f0-5f1 & 5f0-2f4 & 4f2-3f3 & 2f3-1f4\\
         8f0-8f1 & 7f0-4f4 & 5f2-2f3 & 2f3-3f4\\
         9f0-7f1 & 8f0-6f4 & 5f2-3f3 & 3f3-1f4\\
         9f0-8f1 & 1f1-1f2 & 5f2-4f3 & 4f3-4f4\\
         1f0-2f2 & 1f1-2f2 & 5f2-5f3 & 5f3-4f4\\
         1f0-3f2 & 3f1-3f2 & 6f2-4f3 & 5f3-5f4\\
         4f0-3f2 & 5f1-6f2 & 6f2-5f3 & 6f3-5f4\\
         5f0-3f2 & 6f1-7f2 & 7f2-6f3 & 6f3-6f4\\ 
         5f0-4f2 & 7f1-7f2 & 7f2-7f3 & \\ 
         5f0-5f2 & 8f1-8f2 & 8f2-6f3 &  \\ 
		\hline
	\end{tabular}
\end{table}

\bsp	
\label{lastpage}
\end{document}